 \documentclass{aa}
\usepackage{graphicx}
\usepackage{natbib}
\usepackage{epsfig}
\def\ut#1{\mathop{\vtop{\ialign{##\crcr
     $\hfil\displaystyle{#1}\hfil$\crcr\noalign
     {\kern1pt\nointerlineskip}\hbox{$\hfil\sim\hfil$}\crcr
     \noalign{\kern1pt}}}}}

\def\undersymbol#1#2{\mathop{\vtop{\ialign{##\crcr
     $\hfil\displaystyle{#2}\hfil$\crcr\noalign
     {\kern1pt\nointerlineskip}\hbox{$\hfil#1\hfil$}\crcr
     \noalign{\kern1pt}}}}}
\def\arcsec{^{\prime\prime}}
\def\arcmin{^{\prime}}
\def\degr{^0}
\def\hour{^{\rm h}}
\def\minute{^{\rm m}}
\def\second{^{\rm s}}

\begin{document}

 \title{The puzzling symbiotic X-ray system 4U1700+24}
\author{A.A. Nucita\inst{1,2}, S. Stefanelli\inst{1}, F. De Paolis\inst{1,2},  N. Masetti\inst{3},  G. Ingrosso\inst{1,2}, M. Del Santo\inst{4}, \and L. Manni\inst{1,2}}
\institute{Department of Mathematics and Physics {\it ``E. De Giorgi''}, University of Salento, Via per Arnesano, CP 193, I-73100, 
Lecce, Italy \and INFN, Sez. di Lecce, via per Arnesano, CP 193, I-73100, Lecce, Italy \and INAF -- 
Istituto di Astrofisica Spaziale e Fisica Cosmica di Bologna, via Gobetti 101, I-40129 Bologna, Italy \and INAF/IAPS, via del Fosso del Cavaliere, 100, I-00133, Roma, Italy}
   
\offprints{A.A. Nucita, \email{nucita@le.infn.it}}
\date{Submitted: XXX; Accepted: XXX}


{
  \abstract
{Symbiotic X-ray binaries form a subclass of low-mass X-ray binary systems consisting of a neutron star accreting 
material from a red giant donor star via stellar wind or Roche lobe overflow. Only a few confirmed members are currently known;
4U 1700+24 is a good candidate as it is a relatively bright X-ray object, possibly associated with the late-type star V934 Her.}
{We analysed the archive {\it XMM}-Newton and Swift/XRT observations of 4U 1700+24 in order to have a uniform 
high-energy ($0.3-10$ keV) view of the source. Apart from the 2003, 2010, and 2012 data, publicly available but still unpublished, 
we also took the opportunity to re-analyze a set of {\it XMM}-Newton data acquired in 2002.}
{After reducing the {\it XMM}-Newton and Swift/XRT data with standard methods, we performed a detailed spectral and timing analysis.}
{We confirmed the existence of a red-shifted O VIII Ly-$\alpha$ transition (already observed in the 2002 {\it XMM}-Newton data) 
in the high-resolution spectra collected via the RGS instruments. The red-shift of the line is found in all the analysed observations and, on average, it 
was estimated to be $\simeq 0.009$. We also observed a modulation of the centroid energy of the line on short time scales (a few days) and discuss the 
observations in the framework of different scenarios. If the modulation is due to the gravitational red-shift of the neutron star, it might arise from a 
sudden re-organization of the emitting $X$-ray matter on the scale of a few hundreds of km. Alternatively, we are witnessing a uni-polar jet of matter 
(with typical velocity of $1000-4000$ km s$^{-1}$) possibly emitted by the neutron star in an almost face-on system. The second possibility seems 
to be required by the apparent lack of any modulation in the observed $X$-ray light curve. We also note also that the low-resolution spectra 
(both {\it XMM}-Newton and Swift/XRT in the $0.3-10$ keV band) show the existence of a black body radiation emitted by a region 
(possibly associated with the neutron star polar cap) with typical size from a few tens to hundreds of meters. The size of this spot-like region reduces as the overall luminosity of 4U 1700+24 decreases.}
   {}
  
}

   \authorrunning{Nucita et al.}
   \titlerunning{$XMM$-Newton and Swift observation of 4U1700+24}
   \maketitle

\section{Introduction}

Symbiotic X-ray binaries (SyXBs) form a tiny subclass of 
Galactic low-mass X-ray binaries (LMXBs) characterized by a 
red giant star (generally of spectral type M) which loses 
matter to a compact object, most likely a neutron star 
(NS), via stellar wind, or (less frequently, as in the 
case of GX 1+4; \citealt{chakrabarty1997}) Roche lobe 
overflow. Only seven confirmed members are currently known, 
{while for other candidates like 1RXS J180431.1-273932 (\citealt{nucita2007}) 
follow-up observations allowed us to exclude the SyXB nature} (see 
\citealt{masetti2012} and references therein). However, 
according to stellar population synthesis studies performed 
by \citet{lu2012}, between 100 and 1000 of these objects 
are expected to be in the Galaxy (although one should note that 
half of the SyXBs considered by the authors 
were found to be either spurious or unconfirmed cases).

The SyXB subclass only started gaining some attention from the 
scientific community in the last decade only; however, X-ray 
studies of these sources are still quite sporadic, with only 
a handful of objects having been explored in this spectral 
window (\citealt{masetti2002,masetti2007a,masetti2007b}, \citealt{rea2005},  \citealt{paul2005}, 
\citealt{tiengo}, \citealt{mattana2006}, \citealt{patel2007}, \citealt{corbet2008},
\citealt{marcu2011}, \citealt{gonzalez2012}).

One of these sources is 4U 1700+24. It was discovered (\citealt{cooke1978}, \citealt{forman1978}) as a 
relatively bright X-ray object, with variability on both long-term timescales (months to years; 
\citealt{masetti2002}, \citealt{corbet2008}) and short-term timescales (tens 
to thousands of seconds: \citealt{garcia1983}, \citealt{dalfiume1990}); 
This characteristic suggested that the source might be an accreting system; \citet{garcia1983} 
proposed the brightlate-type star V934 Her as the optical counterpart of 4U 1700+24 
(see also \citealt{gaudenzi1999} and \citealt{masetti2002})
on the basis of its position and the detection of emission
lines in its ultraviolet spectrum. This association was later 
confirmed by \citet{masetti2006} with a {\it Chandra} X-ray 
satellite observation that provided a localization of the source 
with subarcsecond precision.

Periodicity studies of the object's light curve were performed in 
X-rays (\citealt{masetti2002}, \citealt{galloway2002}, \citealt{corbet2008}) and optical (\citealt{hinkle2006}), without finding any 
concluding evidence of either orbital or accretor spin modulation.

X-ray spectroscopy of the source, obtained over the last decades 
(\citealt{garcia1983}, \citealt{dalfiume1990}, \citealt{masetti2002}), 
shows a continuum typical of accreting LMXBs, with a thermal 
component probably originating on or near the accretor and a 
Comptonized emission detected up to 100 keV. In particular, \citet{masetti2002} 
examined the X-ray spectroscopic properties of the 
source using data collected with several satellites over 13 years, 
from 1985 to 1998. After this study, \citet{tiengo} published 
a paper on the X-ray behaviour of 4U 1700+24: the 
authors analysed an observation collected with the {\it XMM}-Newton 
satellite in 2002 and found an emission feature at $\approx$ 0.5 keV
and an emission line at 
$\simeq 0.64$ keV which was possibly
identified as the red-shifted O VIII Ly-$\alpha$ transition. 
\begin{table*}[t!]
\caption[]{Log of the archive {\it XMM}-Newton observations analysed in this paper.}
\vspace{-.3cm}
\hspace{-.5cm}
\scriptsize
\begin{tabular}{llllllllll}
\noalign{\smallskip}
\hline
\hline
OBS. ID& REV     &	NOM. RA&	NOM. DEC&  POS. ANGLE&	       DATE &   START             &  END           &NOM.DUR.&EXP. TIME \\
       &	 &	(deg)  &	(deg)   &   (deg)    &	 (yr-m-d)   &   (h:m:s)           &  (h:m:s)       &(ks)    & (ks)         \\
\hline
0155960601 *&$489$&$256.64370$&$23.97183$&$295.63034$&2002-08-11&15:55:49.0&19:32:02.0&$13$&$-,-,3.2$\\       
0151240301 *&$593$&$256.64385$&$23.97183$&$79.403876$&2003-03-07&01:08:08.0&04:32:20.0&$12$&$-,-,1.3$\\        
0151240201 *&$594$&$256.64385$&$23.97183$&$79.399512$&2003-03-09&01:03:49.0&04:27:57.0&$12$&$-,-,5.5$\\        
0151240401  &$673$&$256.64385$&$23.97183$&$294.30723$&2003-08-13&15:23:12.0&19:40:38.0&$15$&$9.4, 9.2, 7.8$\\       
\noalign{\smallskip}
\hline
\noalign{\smallskip}
\multicolumn{10}{l}{Note: The observations labelled with an asterisk presented pile-up in the EPIC cameras. We were able to correct for the pile-up only the EPIC pn data. }\\ 
\multicolumn{10}{l}{Hence, in these cases, we avoided using the MOS 1 and MOS 2 events in order to prevent spurious effects (see text for details); as a 
consequence, a symbol}\\
\multicolumn{10}{l}{$-$ appears in the last column.}\\ 
\noalign{\smallskip}
\hline
\hline
\noalign{\smallskip}
\end{tabular}
\label{table1}
\end{table*}

No further investigations on the X-ray spectroscopic behaviour of
4U 1700+24 have been performed since then; however, three more 
{\it XMM}-Newton pointings performed in 2003 and seven Swift/XRT observations made 
in 2010 and 2012 are publicly available
but still unpublished. In this paper we present an analysis of these data, 
together with an independent examination of the 2002 observation first reported by 
\citet{tiengo} in order to have a uniform analysis of the whole 
{\it XMM}-Newton and Swift/XRT data sets concerning 4U 1700+24. We found that
the feature observed at $\simeq 0.5$ keV is possibly an artifact due 
primarily to the instrumental oxygen edge (see e.g. \citealt{padilla2013}), while 
we confirm the existence of the $\simeq 0.6$ keV line. Moreover, we found that 
the line evolves in time.

The present work is structured as follows: Sect. 2 reports the observations and the 
data analysis; in Sect. 3, the results from these four pointings are 
presented; whereas Sect. 4 provides a discussion of our results.

\section{{\it XMM}-Newton observation and data reduction}
\label{s:xmm1}
The source 4U1700+24 has been observed several times (see Table \ref{table1} for details) by all the $X$-ray instruments 
(RGS 1 and 2, EPIC-MOS 1 and 2, EPIC-pn; \citealt{jansen2001}, \citealt{struder}, and \citealt{turner}) 
on board the {\it XMM}-Newton satellite.  Here, we report the observation identification number, 
the nominal target coordinates (right ascension and declination), the position angle, the observation date together with the start and end time, the nominal 
duration, and the esposure time after removing the high-energy flares.

The observation raw data files (ODFs) were processed using the {\it XMM}-Science
Analysis System (SAS version $13.0.0$) and with the most updated calibration constituent files. To obtain 
the calibrated low- and high-resolution spectra, we ran the {\it emchain} and {\it epchain} tools for the EPIC cameras products, while the 
{\it rgsproc} pipeline was executed for the RGS 1 and RGS 2 instruments.

We followed the standard analysis recipes described in the \citet{xrps}. In particular, we extracted the light curves above 10 keV for the full MOS and pn cameras. 
We then identified and discarded parts of the observations 
affected by high levels of background activity, by using a threshold of 0.35 and 0.40 counts s$^{-1}$ for the two MOS and pn, respectively. 
For each observation, the exposure time resulting from this procedure is reported (in ks) in the 
last column of Table \ref{table1} for the MOS 1, 2 and pn, respectively. The events collected during the 
good time intervals were only used in the spectral analysis, being the timing analysis performed without 
applying any time filters to avoid the introduction of artifacts.
 
The $X$-ray emission from the source was first extracted from a circular region centred on the nominal position of 4U 1700+24
as determined by \citet{masetti2002} when analyzing the {\it Chandra}/HRC data 
($\alpha$= $17\hour$ $06\minute$ $34.517\second$, $\delta$= $+23\degr $ $58\arcmin$ $18.66\arcsec$, with errors on both coordinates of $\simeq 0.6''$) 
and with a radius chosen to contain at least $80\%$ of the total source energy. The background signal was accumulated from circular regions on the same chip. 

We noted that the observations labeled as $0155960601$, $0151240301$, and $0151240201$ were affected by pile-up. This effect consists in 
two or more photons hitting nearby CCD pixels during an exposure, thus producing an event which mimics a single, larger energy photon.  
If not severe, the pile-up effect can be mitigated by following the method described in the \citet{xrps}. In particular, we extracted the source 
signal by using an annular region centred on the source nominal coordinates with inner and outer radii of $\simeq 10''$ and $\simeq 40''$, respectively. 
With this choice, and in accordance with the signal found by \citet{tiengo} when analyzing the 2002 observation, we were able to correct the pile-up 
for the EPIC pn data, but not for the MOS 1 and 2 cameras for which the correction would have  greatly reduced
the number of good counts. Consequently, we avoided using these data to prevent spurious 
effects in the spectral and timing analysis. 

Finally, source and background $X$-ray spectra, together with the associated ancillary and response matrix files, 
were extracted and imported within XSPEC (\citealt{arnaud}) for a simultaneous fitting procedure. 

\section{Swift/XRT observation and data reduction}
\label{s:swift}

Swift/XRT observed 4U 1700+24 in 2010 with two dedicated pointings and in 2012 with 
five observations, because the source was in the same field of view as the gamma ray 
burst GRB121202 A (see Table \ref{tableswift} for details on the archive data sets).

The Swift data were analysed using standard procedures \citep{burrows} and the latest calibration files available at 
$\rm http://heasarc.nasa.gov/docs7swift/analysis/$. In particular, we processed the XRT products  
with the {\it xrtpipeline} (v.0.12.6) tasks, applied standard screening criteria by using ftools (Heasoft v.6.13.0), and extracted with 
{\it xselect} the source spectra and light curves (in the 0.3-10 keV band) from a circular region (with radius of $\simeq 40\arcsec$) 
centred on the nominal coordinates of the target. When possible, the background spectra and light curves were 
also extracted from circular regions. We noted that the 2012 Swift/XRT observations were affected 
by pile-up because the corresponding source count rates were above $\simeq 0.5$ count s$^{-1}$. Thus, we followed the recipe presented in the Swift on-line 
threads\footnote{See $\rm http://www.swift.ac.uk/analysis/xrt/pileup.php$} and discarded the central part (up to $\simeq 15\arcsec$) of the source extraction 
region until the source count rate was below the threshold value. 
We then used the {\it xrtmkarf} task to create the ancillary response files and took into 
account the corrections for the different extraction areas of the source and background and for
vignetting. Finally, the light curves were background corrected.
\begin{table}[t!]
\caption[]{Log of the archive Swift/XRT observations analysed in this paper.}
\vspace{-.3cm}
\hspace{0.0cm}
\scriptsize
\begin{tabular}{lllll}
\noalign{\smallskip}
\hline
\hline
OBS. ID&  DATE      &   START   &  END          &EXP. TIME \\
       & (yr-m-d)   &   (h:m:s) &  (h:m:s)      & (ks)         \\
\hline
0009080001   &2010-05-30& 20:29:06&23:39:56&1.4 \\       
0009080002   &2010-06-04& 12:55:00&00:16:58&9.7 \\        
0054025500 * &2012-12-02& 05:07:33&13:12:24&12.8\\        
0054025501 * &2012-12-02& 13:13:56&23:14:27&6.4 \\        
0054025502 * &2012-12-03& 14:56:26&23:03:44&5.3 \\       
0054025503 * &2012-12-04& 00:47:17&10:28:36&5.8 \\       
0054025504 * &2012-12-05& 13:21:41&18:11:36&5.9 \\ 
\noalign{\smallskip}
\hline
\noalign{\smallskip}
\multicolumn{5}{l}{Note: The observations labeled with an asterisk presented pile-up.}\\ 
\noalign{\smallskip}
\hline
\hline
\noalign{\smallskip}
\end{tabular}
\label{tableswift}
\end{table}

\section{Results}
\subsection{{\it XMM}-Newton RGS spectral analysis of 4U1700+24}
\label{s:result2}
Our study of the {\it X}-ray properties of 4U1700+24 started with the analysis of the first order-spectra 
obtained by the {\it XMM}-Newton gratings. The spectral resolution of RGS in the first-order spectrum 
is FWHM=72 m\AA~and the calibration in wavelength is accurate up to 8 m\AA~, corresponding 
to FWHM$\simeq 620$ km s$^{-1}$ and $\Delta v\simeq 69$ km s$^{-1}$ at 35 \AA ~(\citealt{xmmuserbook}). In the following, 
we use the unbinned RGS 1 and RGS 2 spectra for the quantitative analysis and searched for emission lines.

We then imported the spectra (and the associated response matrices) within XSPEC and simultaneously fit the data. In this respect,
the phenomenological spectral analysis followed a local fit method\footnote{Although in a different context, the {\it local fit method} is described in 
\citet{guainazzibianchi2007}.}, i.e. the unbinned spectra were first divided 
in intervals of $\simeq 100$ channels wide and then Gaussian profiles were added to account for all 
identified emission lines. In this procedure, the line energy, as well as its width and normalization were considered 
as free parameters of the fit. In addition, the local continuum was modelled as a power law with a fixed photon index $\Gamma = 1$ and free normalization. 
In the fit procedure, consistently with \citet{tiengo},  we fixed the column density of Galactic neutral hydrogen to the average value\footnote{We used the 
on-line $N_H$ calculator (available at http://heasarc.nasa.gov/cgi-bin/Tools/w3nh/w3nh.pl) to get the average column density.} 
observed along the line of sight towards 4U1700+24, i.e. $4.4\times10^{20}$ cm$^{-2}$ (\citealt{dickey}).
For line doublets and triplets and for emission lines close to free-bound transitions, the relative distance 
among the central energies was frozen to the value predicted by the atomic physics. 

We used the C-statistics (\citealt{cash}) as the estimator of the goodness of the performed fit and, for any line initially 
recognized as such by eye, the feature was considered {\it detected} if, repeating the fit without any Gaussian profile, the newly obtained value of the
statistics differed by at least 2.3 from the previous one. This choice corresponds to 68$\%$ confidence level (or, equivalently, 
1 $\sigma$ for one interesting parameter) for the detected emission line \citep{arnaud}. Finally, our estimate of the error associated 
with the centroid energy was evaluated as the quadratic sum of the error in output from XSPEC and the calibration uncertainty quoted above. 
\begin{figure}[htbp]
\vspace{7.cm} \includegraphics{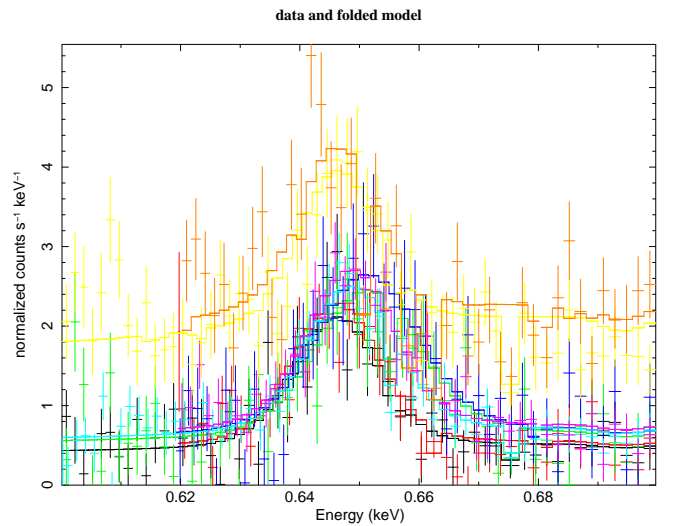}
\caption{A zoom of the RGS 1 and 2 spectra of 4U1700+24 around the O VIII Ly-$\alpha$ emission line (at  about $\simeq 19$ \AA~) for all the observations quoted in Table \ref{table1}. 
The spectra were binned in order to have a signal-to-noise ratio of 5 in each bin, and we give the horizontal axis in energy instead of wavelength. The data in yellow/orange (corresponding to the 
2002 observation, i.e. $0155960601$) clearly show a large continuum component with respect to the other data sets, thus reflecting the high activity of the
source at that time. The solid lines 
represent the best-fit model as described in the text.}
\label{f1}
\end{figure}

In particular, this procedure resulted in the identification of an emission feature at $\simeq 0.645$ keV ($\simeq 19$ \AA~), i.e. consistent 
with that already observed by \citet{tiengo} in the RGS 1 and 2 spectra of the 2002 observation. We do still identify the same line in all 
the subsequent {\it XMM}-Newton observations considered in this work. In Fig. \ref{f1}, we give a zoom of the 
RGS 1 and 2 spectra around the identified emission line (binned in order to have $5\sigma$ per bin, and with the horizontal axis in energy) of 4U1700+24, 
for all the observations quoted in Table \ref{table1}. We note that the data in yellow/orange correspond to the 2002 observation 
(Obs. ID $0155960601$) and clearly show a continuum component larger than that present in the other data sets; this reflects the high 
activity of the source at that time (see also next paragraph).

Some emission lines within a few $\sigma$ of the observed feature are found in the CHIANTI database \citep{chianti}, with the Ly-$\epsilon$ line of 
the $N VII$ having a centroid energy position compatible with that observed in our spectra. However, as first discussed by \citet{tiengo}, 
the probability of the occurrence of this transition is very low, thus pushing toward the interpretation of this line as the 
red-shifted Ly-$\alpha$ transition of the H-like oxygen, i.e. the OVIII feature with a rest-frame wavelength of $18.9671$\AA ~.  

In Table \ref{table2}, we give the best-fit parameters ($C_{Stat}=2234.33$ for 2044 degrees of freedom) 
obtained via the method described above\footnote{When repeating the fit without the Gaussian line profile, the new value of the statistics is
$C_{Stat}=3270.75$ for 2056 degrees of freedom, thus implying the necessity of a Gaussian component.} 
for the emission line observed at $\simeq 19$ \AA ~ for each of the {\it XMM}-Newton observations analysed in this work. Here, we report the mean time of
the observation (in Modified Julian Date, MJD), the line normalization, the best-fit central energy ($E_{obs}$) and the line width ($\sigma_{obs}$).
Assuming that the observed emission line is the red-shifted Ly-$\alpha$ transition of the H-like oxygen (OVIII), in the last two columns we give the 
red-shift factor and associated equivalent velocity (defined as $\Delta \lambda/\lambda=v/c$, being $c$ the speed of light), respectively. 

\begin{figure}[htbp]
\vspace{8.5cm} \includegraphics{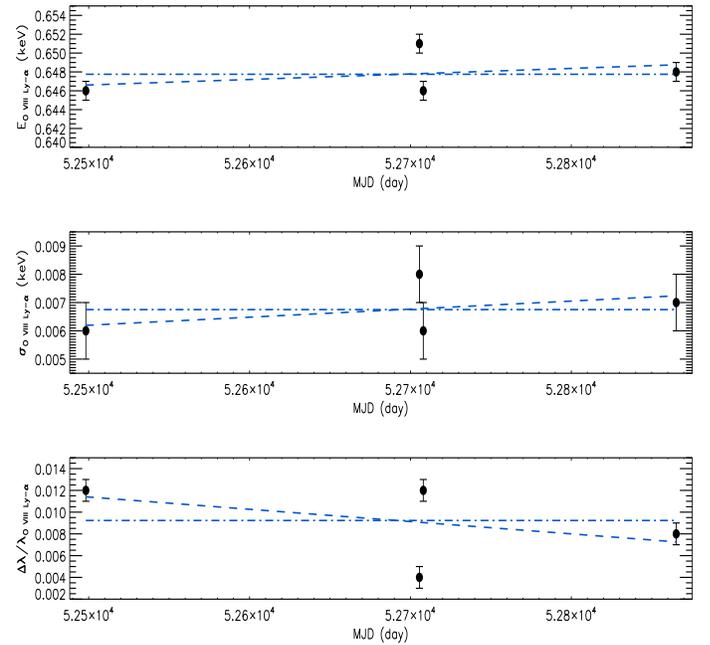}
\caption{Here, we show the centroid position in keV (upper panel) as determined by the Gaussian line fit, 
the line width (middle panel), and the red-shift factor of the O VIII Ly-$\alpha$ line 
identified in the high-resolution spectra of 4U1700+24 as a function of the observation time (defined as the average time of each exposure) in modified Julian date. 
In each panel, the dot-dashed line represents a fit with a constant, while the 
dashed line accounts for any linear trend possibly associated with the data (see text for details).}
\label{fig2}
\end{figure}

On average, the emission line has an energy of $\simeq 0.648$ keV, corresponding to a
wavelength of $\simeq 19.136$\AA ~, and an average red-shift of $\simeq 0.009$ (see also below). 
In Figure \ref{fig2}, we show the position (upper panel) in keV, the line width (middle panel), and the red-shift factor (bottom panel) of the O VIII Ly-$\alpha$ line 
as a function of the observation time (here defined as the average time of each exposure) in MJD. 
We note that the line energy for observation $0155960601$ is consistent with that already
derived by \citet{tiengo}. We first fitted the central energy 
of the emission feature, its width, and the red-shift factor (assuming the rest-frame energy to be that of the O VIII Ly-$\alpha$ line) 
with constant values (see the dot-dashed lines in the three panels in Fig. \ref{fig2}) finding $0.648\pm 0.001$ keV ($\chi^2=2.8$ for 3 dof), 
$0.0067\pm 0.001$ keV ($\chi^2=0.45$ for 3 dof), and $0.009\pm 0.001$ ($\chi^2=2.1$ for 3 dof), respectively. Then, we 
searched for a {global} trend in the data by using a linear model as a fitting function. 
Assuming that the {overall behaviour of the} data varies linearly with time, we obtained the rate of changes of the interested parameters to be 
$(5.9\pm5.4)\times 10^{-6}$ keV day$^{-1}$ ($\chi^2=3.6$ for 2 dof), $(2.8\pm 5.4)\times 10^{-6}$ keV day$^{-1}$ ($\chi^2=0.5$ for 2 dof), 
and $(-1.1\pm0.8)\times 10^{-5}$ day$^{-1}$ ($\chi^2=2.1$ for 2 dof), respectively. Although a close inspection of Fig. \ref{fig2}  might allow us 
to conclude that the emission line position changes {on time scales of $\simeq 2$ days (see e.g. the bottom panel where this effect is amplified by the fact 
that the red-shift factor is given and also the log of the observations in Table \ref{table1}), a constant model seems to be preferred {when one considers the long-term trend}.}
\begin{table*}[t!]
\caption[]{For each of the {\it XMM}-Newton observations, we give the main best-fit parameters (normalization, centroid position energy $E_{obs}$, line width $\sigma_{obs}$, and 
wavelength $\lambda_{obs}$) of the emission feature observed at $\simeq 19$ \AA ~(see text for details).} 
\scriptsize
\begin{center}
\begin{tabular}{llllllll}
\noalign{\smallskip}    
\hline
\hline
OBS. ID         &  MJD        & $N_2$  & ${\rm E_{obs}}$      & ${\rm\sigma_{obs}}$  & ${\rm\lambda_{obs}}$       &${\rm \Delta \lambda / \lambda}$ & ${\rm v}$          \\
                & (day)	      & ${\rm (cm^{-2} s^{-1})}$      & (keV)                & (keV)               & (\AA ~)                    &	                           & km s$^{-1}$        \\
\hline 
$0155960601 $   & $52498.21$&  $(9.0\pm1.0)\times10^{-4}$  & ${0.646^{+0.001}_{-0.001}}$ &$0.006_{-0.001}^{+0.001}$ &${19.195^{+0.030}_{-0.030}}$  & ${0.012^{+0.001}_{-0.002}}$ & ${3600^{+300}_{-600}}$\\
$0151240301 $   & $52705.58$&  $(1.1\pm0.2)\times10^{-3}$  & ${0.651^{+0.001}_{-0.001}}$ &$0.008_{-0.001}^{+0.001}$ &${19.048^{+0.030}_{-0.030}}$  & ${0.004^{+0.002}_{-0.002}}$ & ${1200^{+600}_{-600}}$\\
$0151240201 $   & $52707.99$&  $(8.4\pm1.0)\times10^{-4}$  & ${0.646^{+0.001}_{-0.001}}$ &$0.006_{-0.001}^{+0.001}$ &${19.195^{+0.030}_{-0.030}}$  & ${0.012^{+0.002}_{-0.002}}$ & ${3600^{+600}_{-600}}$\\
$0151240401 $   & $52865.22$&  $(9.9\pm1.0)\times10^{-4}$  & ${0.648^{+0.001}_{-0.001}}$ &$0.007_{-0.001}^{+0.001}$ &${19.136^{+0.030}_{-0.030}}$  & ${0.009^{+0.001}_{-0.002}}$ & ${2400^{+300}_{-600}}$\\
\noalign{\smallskip}                                                                    
\hline
\hline
\noalign{\smallskip}
\end{tabular}
\end{center}
\label{table2}
\end{table*}

We also tried the identification of the most intense lines of He-like ions of oxygen in the range 5-35 \AA, i.e. the transitions between 
the $n=2$ shell and the $n=1$ ground state shell as the resonance line, ${\bf r:~}~1s^2$ $^1S_0$-$1s2p$ $^1P_1$, 
the two inter-combination lines (often blended), ${\bf i:~}~1s$$^2$$^1S_0$-$1s2p$ $^3P_{2,1}$, and the forbidden feature, 
${\bf f:~}$~$1s^2$$^1S_0$-$1s2s$$^3S_1$. These transitions are particularly important since, as demonstrated by \citet{porquet}, their relative 
emission strengths are good indicators of the physical conditions of the gas density and temperature. 

\begin{figure}[htbp]
\vspace{7.0cm} \includegraphics{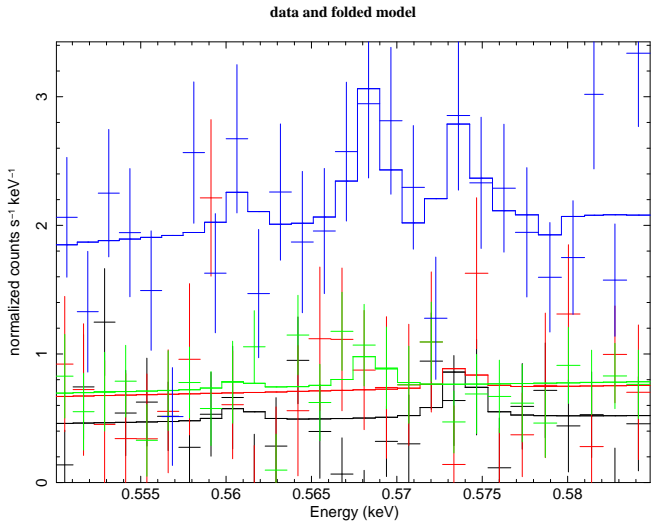}
\caption{A zoom around the O VII triplet (data points rebinned in order to have a signal-to-noise ratio of 5 in each bin) and the best-fit model 
superimposed after red-shifting the rest-frame energy of the complex by $\Delta \lambda/\lambda \simeq 0.009$, i.e. the energy found when analyzing the 
O VIII Ly-$\alpha$.}
\label{fig3}
\end{figure}

Because of the poor statistics of the RGS 
data, the fit the O VII complex with a model constituted by a power law and three Gaussians (with all the parameters free, except the relative distances 
among the lines, as well as the continuum power law index) did not converge. We then used a different approach and, in particular, we fixed
the centroid energy of the interested lines to that expected by the atomic physics, after correcting for the average red-shift ($\Delta \lambda / \lambda 
\simeq 0.009$) previously found. We also set all the line widths to zero. Leaving as free parameters the Gaussian line and power law 
normalizations allowed us to get a reasonable fit (see Fig. \ref{fig3}) characterized\footnote{Performing the fit procedure with a normalized power law only 
resulted in a best-fit with $C_{Stat}=673.0$ for 532 degrees of freedom. In the framework of the local method discussed in the text, the 
comparison of this best-fit with the previous one allows us to be confident with the existence of the O VII complex.} by $C_{Stat}=664.3$ for 520 degrees of freedom. 
However, this procedure only resulted in upper limits to all the line normalizations, thus making impossible to infer the physical 
condition of the X-ray emitting gas. In this case, our best-fit model works only as a guide for the eye when searching for the O VII complex lines. 
In this respect, we note 
that the RGS data show the existence (although with a signal-to-noise ratio less than $\simeq 1$) of emission lines in the positions where the O VII complex lines are expected;
this makes us confident that the line identified at $\simeq 0.19$ \AA ~is really the O VIII Ly-$\alpha$ transition  red-shifted by an average red-shift of $\simeq 0.009$.
\subsection{{\it XMM}-Newton EPIC spectral analysis of 4U1700+24}
\label{s:result3}
Following \citet{masetti2002} and \citet{tiengo}, we simultaneously fitted all the low-resolution spectra of 
4U1700+24 with an absorbed black body plus Comptonization (COMPST in XSPEC). As in the case of the RGS analysis, we fixed the Galactic neutral hydrogen column 
density to the average value observed towards the target, i.e. $4.4\times10^{20}$ cm$^{-2}$ (\citealt{dickey}). However, we noted the existence of large residuals 
at low energies (close to $\simeq 0.5$ keV and $\simeq 0.6$ keV) particularly similar to the typical shape of emission lines. 
The existence of a broad emission ($\sigma \simeq 0.1$ keV) line at $\simeq 0.5$ keV was already noted by \citet{tiengo} and explained 
as due to the blending of several lines possibly observed in the RGS data, but no other emission line feature was reported.
We note that the feature observed at $\simeq 0.5$ keV may be due to the instrumental absorption oxygen edge at $23$ \AA ~as
recently found by \citet{padilla2013}. For simplicity, we model this feature as a large Gaussian line after verifying that the use of 
the model EDGE in XSPEC to account for its shape did not dramatically improve our best-fit.

\begin{figure*}[htbp]
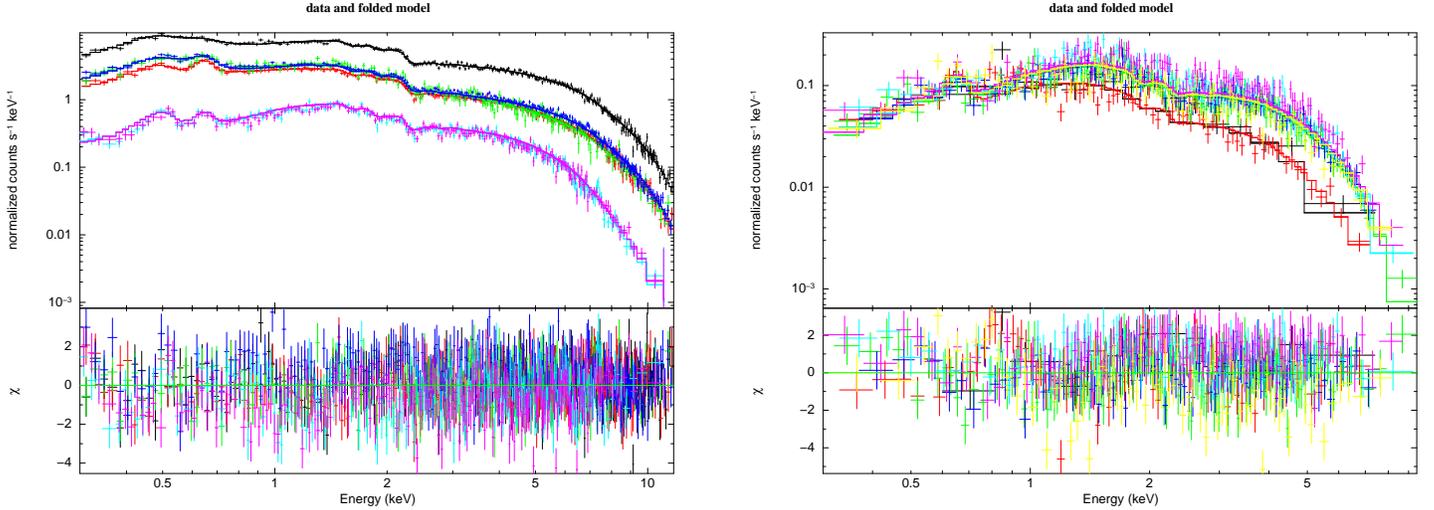

\vspace{7.cm} 
\begin{center}$
\begin{array}{cc}
\includegraphics{best_fit.eps} &
 \includegraphics{swift_bestfit.ps}
\end{array}$
\end{center}
\caption{Left panel: the EPIC spectra of 4U1700+24 during the four observations analysed in this work. Black, green, blue, and red data points correspond to the EPIC pn data of the observations
0155960601, 0151240301, 0151240201, and 0151240401, respectively. Since the pile-up affected most of the data sets, MOS 1 and MOS 2 data 
(purple and cyan data points) were only available for the last observation (i.e. 0151240401). Here, the solid lines correspond to the best-fit model described in the text. 
Right panel: the Swift/XRT spectra of 4U1700+24 during the 2010 and 2012 observations (data points) together with the best-fit model (solid lines). 
Here, the black and red data correspond to the Swift observations 0009080001 and 0009080002, respectively while the other data sets shown (purple, blue, yellow, green and cyan) 
correspond to the observation IDs from 0054025500 to 0054025504 of Table \ref{table3}.
}
\label{bestfitspectra}
\end{figure*}

The existence of something missing in the model at $\simeq 0.6$ keV is particularly clear in the residuals associated to the 2003 EPIC spectra;
we interpret this feature as the fingerprint of the O VIII Ly-$\alpha$ observed in the RGS data. Thus, our best-fit model consisted of an absorbed 
black body plus a Comptonization to which two emission lines (the instrumental feature at $\simeq 0.5$ keV and a real emission line at $\simeq 0.6$ keV) were 
added. For convergence purposes, we left the centroid energy (and the associated width) of the broad feature observed at $\simeq 0.5$ keV as free parameters, but 
fixed the position and width of the O VIII Ly-$\alpha$ line to the value obtained when analyzing RGS 1 and 2 spectra (see Table \ref{table2}). 
\begin{table*}[t!]
\caption[]{The best fit model describing the EPIC data (first four rows) and Swift/XRT data (last seven rows) consists in
an absorbed black body plus a Comptonization to which two emission lines (at $\simeq 0.5$ keV and $\simeq 0.6$ keV) were 
added (see text for details). 
}
\scriptsize
\begin{center}
\begin{tabular}{llllllllll}
\noalign{\smallskip}    
\hline
\hline
ObsID        & $kT_{BB}$       & $N_{BB}$                 & $kT_c$        & $\tau$        & $N_C$              & $E_{1}$& $\sigma_{1}$ & $N_{1}$                                          & $N_{2}$\\
             & (keV)           & $ {\rm R_{\rm km}^2/D_{10}^2}$ & (keV)         &               & ${\rm Nf/4\pi D^2}$       & (keV)     & (keV)           & ${\rm (cm^{-2} s^{-1})}$            & ${\rm (cm^{-2} s^{-1})}$            \\
\hline
0155960601   & ${0.90^{+0.09}_{-0.06}}$ & ${9.00^{+1.16}_{-2.33}}$ & ${2.11^{+0.08}_{-0.06}}$ & ${34.56^{+0.93}_{-2.20}}$ & ${0.0183^{+0.0007}_{-0.0004}}$ & ${0.464^{+0.028}_{-0.011}}$ & ${0.131^{+0.009}_{-0.024}}$ & ${0.017^{+0.001}_{-0.003}}$     & $\le{0.0004}$ \\
0151240301   & ${0.82^{+0.07}_{-0.03}}$ & ${6.94^{+1.76}_{-1.39}}$ & ${2.12^{+0.20}_{-0.09}}$ & ${27.93^{+2.68}_{-2.68}}$ & ${0.0072^{+0.0005}_{-0.0006}}$ & ${0.485^{+0.020}_{-0.028}}$ & ${0.093^{+0.035}_{-0.024}}$ & ${0.005^{+0.002}_{-0.001}}$     & ${0.0007^{+0.0005}_{-0.0004}}$ \\
0151240201   & ${1.13^{+0.10}_{-0.10}}$ & ${2.66^{+0.24}_{-0.26}}$ & ${2.45^{+0.45}_{-0.22}}$ & ${24.00^{+1.88}_{-3.60}}$ & ${0.0071^{+0.0003}_{-0.0002}}$ & ${0.490^{+0.012}_{-0.011}}$ & ${0.043^{+0.014}_{-0.012}}$ & ${0.0019^{+0.0004}_{-0.0004}}$  & ${0.0010^{+0.0001}_{-0.0001}}$ \\
0151240401   & ${0.71^{+0.05}_{-0.05}}$ & ${6.98^{+2.25}_{-1.41}}$ & ${1.96^{+0.05}_{-0.05}}$ & ${33.07^{+1.88}_{-1.27}}$ & ${0.0075^{+0.0002}_{-0.0004}}$ & ${0.487^{+0.008}_{-0.022}}$ & ${0.108^{+0.018}_{-0.010}}$ & ${0.007^{+0.001}_{-0.001}}$     & ${0.0007^{+0.0001}_{-0.0003}}$ \\
\hline
\hline
0009080001 & ${1.02^{+0.11}_{-0.08}}$ & ${0.50^{+0.10}_{-0.10}}$ & $(2.16)$ & $16.29^{+3.44}_{-2.89}$ & ${0.0008^{+0.0001}_{-0.0001}}$ & -- & -- & --     & ${0.0003^{+0.0001}_{-0.0001}}$ \\
0009080002 &                          &                          &                                &                                         &    &    &        &                                \\
0054025500 & ${0.89^{+0.19}_{-0.13}}$ & ${5.02^{+3.20}_{-1.96}}$ & $(2.16)$ & $28.96^{+2.88}_{-3.10}$ & ${0.0070^{+0.0008}_{-0.0009}}$ & -- & -- & --     & ${0.0012^{+0.0005}_{-0.0005}}$ \\
0054025501 & ${1.05^{+0.28}_{-0.23}}$ & ${2.3^{+2.0}_{-1.0}}$ & $(2.16)$ & $27.34^{+5.34}_{-9.70}$ & ${0.0040^{+0.0008}_{-0.0005}}$ & -- & -- & --     & ${0.0009^{+0.0005}_{-0.0001}}$ \\
0054025502 & ${0.77^{+0.17}_{-0.10}}$ & ${5.6^{+4.1}_{-2.6}}$ & $(2.16)$ & $32.58^{+7.90}_{-4.90}$ & ${0.0031^{+0.0008}_{-0.0005}}$ & -- & -- & --     & ${0.0011^{+0.0005}_{-0.0005}}$ \\
0054025503 & ${0.98^{+0.30}_{-0.16}}$ & ${4.7^{+3.5}_{-1.9}}$ & $(2.16)$ & $31.94^{+9.30}_{-3.90}$ & ${0.004^{+0.001}_{-0.001}}$ & -- & -- & --     & ${0.0013^{+0.0006}_{-0.0006}}$ \\
0054025504 & ${0.88^{+0.29}_{-0.17}}$ & ${1.4^{+1.4}_{-0.7}}$ & $(2.16)$ & $28.90^{+4.28}_{-4.80}$ & ${0.0020^{+0.0003}_{-0.0003}}$ & -- & -- & --     & ${0.0011^{+0.0003}_{-0.0003}}$ \\
\noalign{\smallskip}
\hline                                                                                                                                                                                                                                                                                
\hline
\noalign{\smallskip}
\end{tabular}
 \end{center}
\label{table3}
\end{table*}
With reference to Fig. \ref{bestfitspectra} (left panel), the black, green, blue, and red data points correspond to the EPIC pn data of the observations
0155960601, 0151240301, 0151240201, and 0151240401, respectively. The pile-up affected most of the data sets, so that the MOS 1 and MOS 2 data (purple and cyan points in the same figure) were 
only available for the last observation. The EPIC best-fit parameters ($\chi^2=1.3$ for 1379 degree of freedom - d.o.f) 
are reported in the first four rows of Table \ref{table3}. Here, $kT_{BB}$ is the temperature of the black-body component, 
$kT_{c}$ and $\tau$ the temperature and optical depth of the Comptonization, $E_1$ and $\sigma_1$ the broad feature (at $\simeq 0.5$ keV) 
position and line width, respectively. All the normalizations  of the model components are free to vary. In particular, $N_{BB}$, $N_C$, $N_1$ and $N_2$ are 
the black body, the Comptonization, the broad feature and the O VIII Ly-$\alpha$ line normalizations, respectively.   
In the BBODYRAD normalization, $R_{\rm km}$ and $D_{10}$ are the source radius (in units of km) and distance (in units of 10 kpc), 
respectively. For the COMPST normalization, $N$ represents the total number of photons from the source and $f$ a factor depending on 
the injected photon energy and spectral index.

In Table \ref{table4}, we also give the flux in the 
same band for each of the {\it XMM}-Newton observations (first four rows) and the estimated fluxes in the energy bands $0.3-2$ keV, $2-10$ keV, and $0.3-10$ keV , respectively.
We note that the absorbed 0.3-10 keV band flux results in $2.35\times 10^{-10}$ erg cm $^{-2}$ s $^{-1}$ when averaged over the four {\it XMM}-Newton observations. 
As can be seen in the table, the source was 
in high state during the 2002 observation (see also \citealt{tiengo}), because the flux was a factor of 2 larger than the average value.  
We also note that archival EXOSAT, ROSAT, ASCA, RXTE, and BeppoSAX observations (spanning the years 1985-1998) have shown that the $X$-ray emission 
from the source appears to become harder as its luminosity increases (see e.g. Table 2 in \citealt{masetti2002}). However, we did not find 
this behaviour in the {\it XMM}-Newton data analysed in the present paper. We also note that, based on a set of 
{\it Chandra}/HRC observations, the source again appeared in a high state in April 2005 (with a 2-10 keV flux of 
$\simeq 2\times 10^{-10}$ erg cm $^{-2}$ s $^{-1}$ corresponding to a luminosity of $\simeq 4\times 10^{33}$ erg cm s $^{-1}$, \citealt{masetti2006}).

We also note that the black-body component required by the best-fit model allowed us to estimate the radius of the $X$-ray emitting region to be 
$\simeq 30-130$ m, i.e. consistent with the expected size of a polar cap emission in a NS. 

\begin{table*}[t!]
\caption[]{The $0.3-2$ keV, $2-10$ keV and $0.3-10$ keV band 
fluxes together with the estimated luminosity (full band) for a source distance of $\simeq 420$ pc.}
\scriptsize
\begin{center}
\begin{tabular}{llllll}
\noalign{\smallskip}    
\hline
\hline
ObsID      & MJD        &$F_{{\rm 0.3-2.0 kev}}$                    &$F_{{\rm 2-10.0 kev}}$                     & $F_{{\rm 0.3-10.0 kev}}$       & $L_{{\rm 0.3-10.0 kev}}$      \\
           & (day)      &($10^{-10}$ erg cm $^{-2}$ s $^{-1}$)      &($10^{-10}$ erg cm $^{-2}$ s $^{-1}$)      & ($10^{-10}$ erg cm $^{-2}$ s $^{-1}$) & ($10^{34}$ erg s $^{-1}$)\\
\hline
0155960601 & 52498.22 &${0.669^{+0.006}_{-0.021}}$ &${4.35^{+0.02}_{-0.12}}$ & ${5.02^{+0.03}_{-0.17}}$ & ${1.060^{+0.010}_{-0.040}}$    \\
0151240301 & 52705.58 &${0.290^{+0.005}_{-0.021}}$ &${1.23^{+0.02}_{-0.19}}$ & ${1.52^{+0.03}_{-0.16}}$ & ${0.320^{+0.010}_{-0.040}}$    \\
0151240201 & 52707.99 &${0.245^{+0.005}_{-0.004}}$ &${1.25^{+0.01}_{-0.09}}$ & ${1.49^{+0.01}_{-0.08}}$ & ${0.310^{+0.002}_{-0.020}}$    \\
0151240401 & 52865.22 &${0.284^{+0.002}_{-0.004}}$ &${1.41^{+0.01}_{-0.03}}$ & ${1.69^{+0.01}_{-0.04}}$ & ${0.355^{+0.002}_{-0.010}}$    \\
\hline
\hline
0009080001 & 55346.92 &${0.031^{+0.014}_{-0.008}}$ &${0.080^{+0.015}_{-0.007}}$ & ${0.110^{+0.020}_{-0.010}}$ & ${0.023^{+0.013}_{-0.006}}$    \\
0009080002 & 55351.78 &                            &                            &                             &                                \\
0054025500 & 56263.38 &${0.246^{+0.007}_{-0.020}}$ &${1.30^{+0.06}_{-0.10}}$ & ${1.54^{+0.06}_{-0.13}}$       & ${0.324^{+0.013}_{-0.027}}$    \\ 
0054025501 & 56263.76 &${0.145^{+0.006}_{-0.016}}$ &${0.773^{+0.093}_{-0.146}}$ & ${0.920^{+0.180}_{-0.160}}$ & ${0.193^{+0.038}_{-0.034}}$    \\ 
0054025502 & 56264.79 &${0.143^{+0.005}_{-0.016}}$ &${0.739^{+0.045}_{-0.142}}$ & ${0.880^{+0.110}_{-0.160}}$ & ${0.185^{+0.023}_{-0.034}}$    \\ 
0054025503 & 56265.23 &${0.186^{+0.007}_{-0.037}}$ &${1.09^{+0.08}_{-0.22}}$ & ${1.27^{+0.10}_{-0.27}}$       & ${0.267^{+0.021}_{-0.057}}$    \\ 
0054025504 & 56266.66 &${0.077^{+0.003}_{-0.012}}$ &${0.373^{+0.029}_{-0.086}}$ & ${0.450^{+0.030}_{-0.093}}$ & ${0.095^{+0.006}_{-0.020}}$    \\ 
\hline                                                                                                                                                                                                                                                                                
\hline                                                                                                                                                                                                                                                                                
\noalign{\smallskip}
\end{tabular}
\end{center} 
\label{table4}
\end{table*}

\subsection{Swift/XRT spectral analysis results}
\label{s:resultswift1}

The 2010 and 2012 Swift/XRT source spectra (as well as the corresponding 
background, ancillary, and response matrix files) were imported within XSPEC, grouping together 
the spectra acquired {in 2010}. We note that when a row is empty, each 
parameter remains unchanged with respect to the previous value. Each spectrum was rebinned with a minimum of 25 counts per energy bin. 
We first tried to use the model providing the best-fit for the {\it XMM}-Newton EPIC data (see the previous section), but soon 
we realized that the Gaussian line component at $\simeq 0.5$ keV is not required. On the contrary, 
residuals appeared at $\simeq 0.6$ keV, thus forcing us to maintain a Gaussian component to account for this line 
feature: in particular, we fixed (as before) the line centroid energy to the average 
value $\simeq 0.646$ keV. In order to avoid bad convergences, we fixed the 
the temperature parameter $kT_c$ of the COMPST component to the average value ($\simeq 2.16$ keV) estimated with the 
{\it XMM}-Newton EPIC data only, while leaving the optical depth $\tau$ and the normalizations of the model components free to vary.

The best-fit procedure ($\chi^2= 1.2$ for 621 d.o.f.) resulted in the parameter values 
reported in the last seven rows of Table \ref{table3} (from which one can note
that the spectral properties of 4U1700+24 did not change substantially) and in the estimates of the 
0.3-2.0 keV, 2.0-10.0 keV, and 0.3-10.0 keV band fluxes (see Table \ref{table4}), respectively. The best-fit model is superimposed 
on the Swift/XRT data in the right panel in Fig. \ref{bestfitspectra}. As one can note, 
in accordance with the {\it XMM}-Newton data, a decrease in the $X$-ray luminosity is always accompanied by a decrease 
in the emitting region size. 
{In particular, the 2010 observations clearly show that 
the black-body component may come from an area with a radius that is 
at least one fourth of that estimated with the 2003 {\it XMM}-Newton data.}

\subsection{{\it XMM}-Newton and Swift/XRT temporal analysis results}
\label{s:result4}
\begin{figure*}[htbp]
\vspace{13.cm} \includegraphics{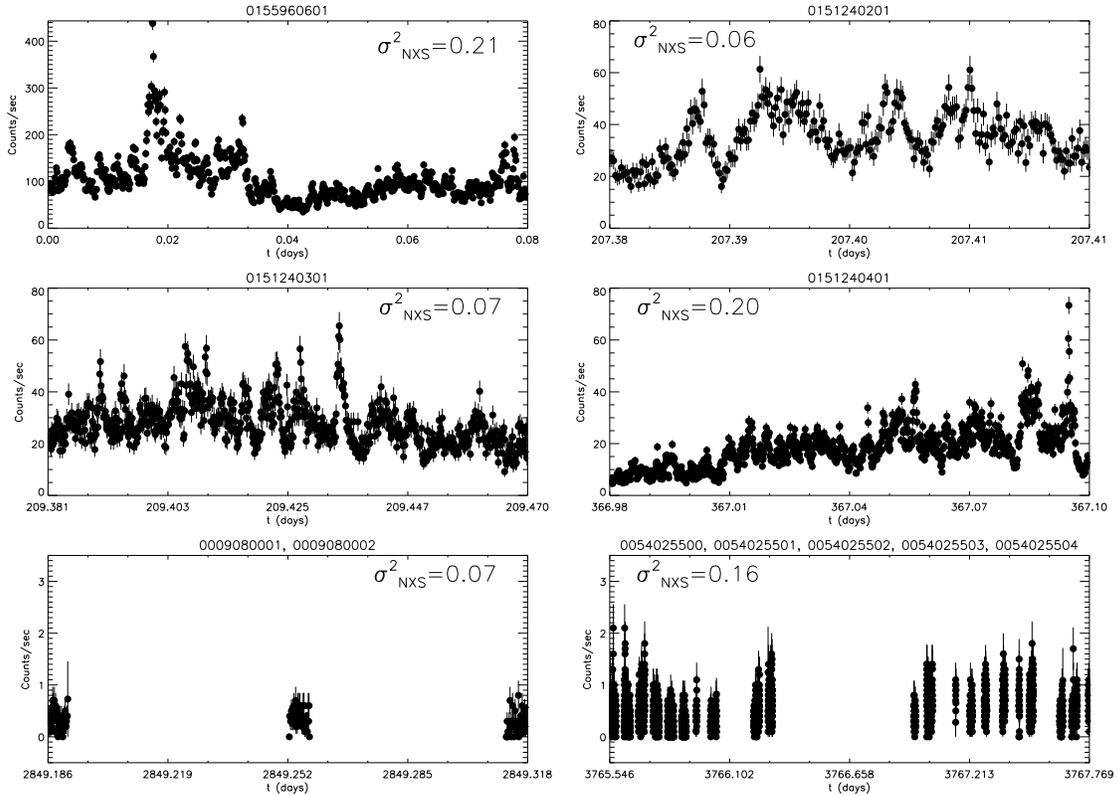}
\caption{The {\it XMM}-Newton and Swift/XRT light curves associated to all the observations analysed in this work 
(see Table \ref{table1}) are shown in separated panels. In each panel, we also give
the estimate of the excess variance of the {\it X}-ray signal during each observation. Time is in days from the beginning of the 2002 {\it XMM}-Newton observation.}
\label{figlc}
\end{figure*}

As discussed before, when extracting the light curve of the target, we did not filter out 
any period of large background activity as this procedure might introduce spurious features in the timing analysis. Hence, 
we used the original event list (corrected for the solar system barycenter when needed) files and extracted the 
light curves in the $0.3-10$ keV energy band (and bin size of 10 seconds) for the source and background. In this procedure, we avoided 
using the data affected by severe pile-up while, for the observation $0151240401$, all the EPIC cameras were used 
to get a final (averaged) light curve.
The obtained (synchronized) light curves were given in input to the SAS task {\it epiclccorr} to account for the background subtraction 
and for the absolute and relative corrections.
The EPIC light curves associated with the four observations analysed in this work (see Table \ref{table1}) are shown in the four first panels of 
Fig. \ref{figlc}. During the four {\it XMM}-Newton observations, the target had an average count rate of $105\pm 7$ count s$^{-1}$, 
$35\pm 4$ count s$^{-1}$, $28\pm 3$ count s$^{-1}$, and $18\pm 2$ count s$^{-1}$, respectively: clearly, as also discussed before, 
the target was characterized by a high state during the 2002 observation.

For 2010 and 2012 Swift/XRT observations, we extracted the source and background light curves (in the 0.3-10 keV energy band and bin-size of 10 sec) 
by using the {\it xselect}, accounting for the pile up when necessary. Since these light curves were also synchronized, 
we subtracted the background and scaled for the extraction areas by using the {\it lcmath} tool. The Swift/XRT light curves are reported 
in the last two panels of Fig. \ref{figlc} with, in particular, the 2010 time series on the left and the 2012 on the right part of the figure, respectively. 
We note that 4U 1700+24 had an average count rate of $0.27\pm 0.23$ count s$^{-1}$ in 2010, and $0.47\pm 0.43$ count s$^{-1}$ in 2012. 

The source is clearly variable and we gave an estimate of this temporal variability by using the normalized excess variance ($\sigma^2_{NXS}$; see e.g. 
\citealt{nandra1997} and \citealt{edelson2002}) to which we associated an uncertainty according to eq. (11) in \citet{vaughan2003}. In each panel in Fig. 
\ref{figlc}, we give the excess variance for the EPIC 0.3-10 keV light curve with a bin-size of 10 seconds. Keeping in mind that 
negative values of $\sigma^2_{NXS}$ indicate the absence of or very small variability in the time series, we conclude that the 4U1700+24 light curve shows 
a certain degree of intrinsic variability which seems to be constant in time. Consistently 
with the results of \citet{masetti2002}, a variability on time-scales of tens to thousands of seconds can be identified in the high-energy light curve. 

We searched for periodic modulations in the time range from 20 s to a few hours by using the the Lomb-Scargle technique 
(\citealt{scargle1982}). We tested the significance of each peak observed in the periodogram by simulating 
5000 simulated red-noise light curves each of which with the same statistical properties (mean, variance, time gaps, and red-noise index) as the observed time series 
following the method described in \citet{timmer}. For each simulated light curve, we evaluated the Lomb-Scargle 
periodogram and calculated the {\it global} probability as explained in \citet{benlloch}. In particular, the global significance 
of a peak at a given frequency and with given amplitude is evaluated by counting the number of peaks 
with the same height (or larger) in the full range of tested frequencies. As a result, we did not detect any clear periodicity with significance larger than 
$\simeq 1 \sigma$.

\section{Discussion and conclusions}
\label{s:resul4}
The large collecting area of the {\it XMM}-Newton telescope allowed us to
study in detail the accreting binary 4U 1700+24. By using the high spectral 
resolution of the RGS instruments, we identified an emission line at $\simeq 0.6$ keV 
(first observed by \citealt{tiengo} in 2002), also present in the pn spectrum of the source. 
We associated the observed emission feature with the O VIII Ly-$\alpha$ transition 
red-shifted by the quantity $\Delta \lambda / \lambda = 0.009\pm 0.001$ as obtained averaging the red-shift factors estimated 
in each single pointing (see Fig. \ref{fig2}).

We also searched in the range $5-35$ \AA ~for the most intense lines of He-like ions of oxygen: the transitions between 
the $n=2$ shell and the $n=1$ ground state shell as the resonance line, ${\bf r:~}~1s^2$ $^1S_0-1s2p$ $^1P_1$, 
the two inter-combination lines (often blended), ${\bf i:~}~1s^2$ $^1S_0-1s2p$ $^3P_{2,1}$, and the forbidden feature, 
${\bf f:~}$~$1s^2$ $^1S_0-1s2s$ $^3S_1$. However, because of the poor statistics of the RGS 
data, a blind fit procedure to the RS data around the O VII complex with a model constituted by a power law and three Gaussians (with all the parameters free, except the relative distances 
among the lines and the continuum power law index) did not converge. Hence, we fixed
the centroid energy of the interested lines to that expected by the atomic physics after correcting for the average red-shift previously found. 
Interestingly, the RGS data show the existence (although with a small signal-to-noise ratio) of emission lines 
in the positions where the O VII complex lines are expected: 
this makes us confident that the line identified at $\simeq 0.19$ \AA ~is the O VIII Ly-$\alpha$ transition 
red-shifted (on average) by $\simeq 0.009$.

A red-shift of the O VIII Ly-$\alpha$ line in the range $0.002-0.013$ (see the estimated values given in Table \ref{table2}) 
can be explained (see also \citealt{tiengo}) as the gravitational red-shift of the photons emitted by a plasma blob at distance $R$ from
an object with mass $M$, i.e.  $\Delta \lambda / \lambda=1/(g_{00})^{0.5}-1$ with $g_{00}=1-2GM/Rc^2$. 
As can be seen, the possibility that 4U 1700+24 hosts a white dwarf 
can easily be ruled out because, for the typical values of white dwarf mass and radius ($M\simeq 1$ M$_{\odot}$ and $R\simeq 2\times 10^4$ km), the expected gravitational 
red-shift is a factor of $10$ (or more) lower than the observed value. In agreement with \citet{garcia1983}, this supports the idea that 4U 1700+24 is a neutron star that 
accretes matter
from a red giant. Assuming a neutron star mass of $\simeq 1.4$ M$_{\odot}$ in the 4U 1700+24 binary system, 
the detected red-shift range corresponds to the gravitational red-shift 
of a photon emitted at a distance of $160-1000$ km from the central object, i.e. consistent with the value found by \citet{tiengo}
when analyzing the 2002 {\it XMM}-Newton observation. Furthermore, a close inspection of Fig. \ref{fig2} allows us to conclude that 
the red-shift of the O VII Ly-$\alpha$ line is variable on a time scale 
of few days (see the log of the observations in Table \ref{table1}). In particular, the red-shifts estimated for the central 
observations 0151240301 and 0151240201 are $\simeq 0.004$ and $\simeq 0.012$, respectively. Since these estimates differ from the average red-shift
value by more than $3-5\sigma$, we are confident that the effect is real. Excluding Doppler contributions due to the orbital motion of any blob of plasma around 
the neutron star (as the associated signatures would be different to the observations presented here), we conclude that we are witnessing the re-organization 
of matter at a distance of a few hundred kms around the accreting object. An alternative picture would be a jet of matter (with typical velocity of 
$1000-4000$ km s$^{-1}$) possibly emitted away by the neutron star in an almost face-on system. The alternative condition seems to be required by 
the apparent lack of any periodicity and/or modulation (as we have verified via a Lomb-Scargle analysis) in the observed $X$-ray light curve. 
However, as also observed by \citet{tiengo}, the puzzling lack of any blue-shifted component implies the necessity of an 
ad-hoc geometry to explain the observations or one could invoke a uni-polar jet emitted by the neutron star.
 
Based on these facts, we prefer a scenario in which the mass coming from the M-type companion stellar wind (see \citealt{postnov2011} and references therein 
for details on the wind accretion in symbiotic $X$-ray binaries) 
is captured directly onto a small zone of the NS surface. The $X$-ray photons emitted are reprocessed by a blob of matter at a few hundred kms from the NS surface 
so that the output emission features are gravitationally red-shifted. 

We estimated the 0.3-10 keV band flux to be $2.35\times 10^{-10}$ erg cm $^{-2}$ s $^{-1}$ when averaged over the four {\it XMM}-Newton observations. 
In Table \ref{table4} we also give the flux in the same band for each of the {\it XMM}-Newton data sets (first four rows) as well as the estimated fluxes 
in the energy bands $0.3-2$ keV and $2-10$ keV, respectively. As is clear, the source was 
in high state during the 2002 observation (see also \citealt{tiengo}), because the flux is a factor of 2 larger than the average value, and seems 
to become fainter with time. This behaviour is also confirmed by the spectral analysis of seven Swift/XRT pointings towards 4U 1700+24 made in 2010 and 2012, when 
the source is a factor of $\simeq 100$ fainter than in 2002. An intermediate luminosity was observed in 2005 (\citealt{masetti2006}). We also observed that a
black-body component is required by the best-fit model. We thus estimated the radius of the $X$-ray emitting region to be a few hundreds of meters, 
which is consistent with the expected size of a polar cap emission in a NS, and observed that this size decreases (down to a few tens of meters) 
as the overall $X$-ray luminosity decreases. 

{In the hypothesis that 4U1700+24 is a symbiotic X-ray binary, we expect it to behave like other candidates of this class of objects, in particular we expect the X-ray light curve 
to show a clear feature at the NS spin period. Typically, the observed period for a symbiotic X-ray system is in the range of minutes to hours (\citealt{lewin1971}, \citealt{chakrabarty1997}, 
\citealt{masetti2007b}, \citealt{corbet2008}, and \citealt{nespoli2010}) but, as already stressed, the timing analysis of the 4U1700+24 light curve in the time range from 20 s to a few hours 
(conducted by requiring at least three full test cycles) did not show any significant feature\footnote{The absence of a clear periodicity 
is not new in these kinds of objects; see e.g. the case of XTE J1743-363 described in \citet{bozzo2013}.}. A much more detailed analysis will be presented elsewhere.

If the source is a member of the X-ray pulsar class, the absence of a periodicity is also expected. 
Considering the X-ray pulsar average properties as given in \citet{Kargaltsev}, 
from their relation $L_{X}=\eta \dot{E}$, where $L_X$ is the 
$0.5-8$ keV luminosity and $\dot{E}$ is the spin-down energy rate (related to 
the NS period $P$ and first period derivative $\dot{P}$ by $\dot{E}\simeq 10^{46} \dot{P}/P^3$ in cgs units), and assuming an average efficiency $\eta = 10^{-4}$, we obtained $\dot{E}\simeq 10^{37}-10^{38}$  erg s$^{-1}$. 
Since the measured $\dot{P}$ values 
are in the range $\sim 10^{-13}-10^{-20}$ s s$^{-1}$ (see the on-line catalogue $\rm http://www.astro.ufl.edu/~anuviswanathan/cgi-bin/psrcat.htm$ and also \citealt{becker2001}), we get $P$ in the range $2\times 10^{-4}-10^{-1}$ s, 
which is clearly not detectable in the {\it XMM}-Newton and Swift data analysed in this work. Interestingly enough, assuming a polar cap model for the wind accreting NS 
(see e.g. \citealt{becker2012}), it is possible to estimate the NS surface 
magnetic field which turns out to be $\ut< 2\times 10^{10}$ G, i.e. in agreement with the typical magnetic field values of the $X$-ray emitting pulsars.

Obviously, a long {\it XMM}-Newton observation or a planned exposure with the pn camera in timing mode could allow us
to infer the NS period. 
}

\begin{acknowledgements}
This paper is based on observations by {\it XMM}-Newton, an
ESA science mission with instruments and contributions directly funded by ESA
Member States and NASA. Part of this work is based on archival data, software or on-line services provided 
by the ASI Science Data Center (ASDC), Italy. 
AAN is grateful to Sara A. A. Nucita for the interesting discussions while preparing this manuscript.
MDS thanks the Department of Mathematics and Physics {\it E. De Giorgi} at the University of Salento and the 
astrophysics group for the hospitality.

\end{acknowledgements}


\end{document}